\documentclass[manuscript]{article}
\usepackage[normalem]{ulem}
\usepackage{graphicx}

\newcommand{\arcsec}{$^{\prime\prime}$}

\begin{document}

\title{A binary main belt comet}

\author{Jessica Agarwal$^{1}$, David Jewitt$^{2,3}$, Max Mutchler$^4$, Harold Weaver$^5$, Stephen Larson$^6$}

\maketitle 

\noindent{\sl 
$^1$Max Planck Institute for Solar System Research, Justus-von-Liebig Weg 3, 37077 G\"ottingen, Germany\\
$^2$Department of Earth, Planetary and Space Sciences,
University of California at Los Angeles, 
595 Charles Young Drive East, 
Los Angeles, CA 90095-1567\\
$^3$Department of Physics and Astronomy,
University of California at Los Angeles,
430 Portola Plaza, Box 951547,
Los Angeles, CA 90095-1547\\
$^4$Space Telescope Science Institute, 3700 San Martin Drive, Baltimore, MD 21218 \\
$^5$The Johns Hopkins University Applied Physics Laboratory, 11100 Johns Hopkins Road, Laurel, Maryland 20723  \\
$^6$Lunar and Planetary Laboratory, University of Arizona, 1629 E. University Blvd., Tucson AZ 85721-0092 \\
}

\noindent{\bf The asteroids are primitive solar system bodies which evolve both collisionally and through disruptions due to rapid rotation \cite{marzari-rossi2011}. These processes can lead to the formation of binary asteroids \cite{pravec-harris2007,walsh-richardson2008,walsh_AIV} and to the release of dust \cite{jewitt_R3-2}, both directly and, in some cases, through uncovering frozen volatiles. In a sub-set of the asteroids called main-belt comets (MBCs), the sublimation of excavated volatiles causes transient comet-like activity \cite{hsieh-jewitt2006,jewitt2012,jewitt_AIV}. Torques exerted by sublimation measurably influence the spin rates of active comets \cite{keller-mottola2015_rotation} and might lead to the splitting of bilobate comet nuclei \cite{hirabayashi-scheeres2016}.
The kilometer-sized main-belt asteroid 288P (300163) showed activity for several months around its perihelion 2011 \cite{hsieh_288P-1}, suspected to be sustained by the sublimation of water ice \cite{licandro_288P} and supported by rapid rotation \cite{agarwal-jewitt2016}, while at least one component rotates slowly with a period of 16\,hours \cite{waniak_288P_dps}.
288P is part of a young family of at least 11 asteroids that formed from a $\sim$10\,km diameter precursor during a shattering collision 7.5 $\times$ 10$^6$ years ago \cite{novakovic-hsieh2012}. 
Here we report that 288P is a binary main-belt comet. 
It is different from the known asteroid binaries for its combination of wide separation, near-equal component size, high eccentricity, and comet-like activity. The observations also provide strong support for sublimation as the driver of activity in 288P and show that sublimation torques may play a significant role in binary orbit evolution.}

Hubble Space Telescope (HST) images from December 2011 revealed  that 288P could be a  binary system at the limits of resolution \cite{agarwal-jewitt2016}. 
Shortly before the next perihelion passage (2016 November 08, at 2.45 AU from the Sun) 288P passed close to Earth (2016 September 11, at 1.45 AU). The proximity to the Earth made it possible to observe 288P with the HST at a spatial resolution sufficient to clearly resolve the two components of the binary (Figure 1 and Extended Data Table 1). The components of 288P lie close to the heliocentric orbital plane (Extended Data Figure 1). The mass of the system, derived from Kepler's third law, is in the range ($1.3\times10^{12} < M <$ 1.1$\times$10$^{13}$) kg, while we cannot meaningfully constrain the density due to the unknown shapes of the components (see Methods).

The two components are similar in their average brightness (Extended Data Table 2), indicating that they are of similar size. 
At the resolution of the data, we cannot determine which component is the source of the dust, or whether both might be. 
With no means to distinguish the two nuclei in the images we instead base our orbit analysis only on the time-dependence of their apparent separation. 
We searched a wide parameter space for binary orbit solutions that reproduce the measured component separations (Fig.~2, see also Methods and Extended Data Figure 2). Orbits having small eccentricities do not fit the data. The only acceptable solutions have eccentricities, $e>$0.6, and fall into three distinct groups characterised by orbital periods near 100, 135, and 175 days, respectively. These groups all have ratios of the orbital semimajor axis to the primary object radius $\sim$100, much larger than the ratios ($<$10) found in most asteroid binaries (Fig.~3). While binary asteroids are common \cite{margot_AIV} 288P is the first to show a wide separation, high eccentricity, similarly sized components and mass-loss activity, suggestive of a different origin.

The HST observations show that 288P re-activated not later than July 2016. Repeated activity near perihelion is a strong indicator of the sublimation of water ice due to increased solar heating. A model of the motion of the dust under the influence of solar gravity and radiation pressure suggests that the activity began with a brief release of comparatively large (millimetre-sized) grains in July, while from mid-September until at least the end of January 2017 (the last of our observations), the dominant grain size fell to $\sim$10\,$\mu$m (Extended Data Figure~3). 
This indicates that the developing gas production first lifted a layer of large, loosely connected grains, possibly deposited around the end of the previous period of activity in 2011/12 \cite{rotundi-sierks2015}. After their removal and with decreasing heliocentric distance, the gas drag became sufficiently strong to lift also smaller particles. The dust production rates were of order 0.04-0.1\,kg\,s$^{-1}$ (see Methods and Extended Data Figure~4), in contrast to 1\,kg\,s$^{-1}$ inferred from 2011 data \cite{agarwal-jewitt2016}.

The  majority of small binary asteroid systems (Fig.~3) likely formed by rotational fission \cite{pravec-harris2007,walsh-richardson2008,walsh_AIV} and subsequently evolved under the action of tides and weak radiation torques. The post-formation evolution depends on the relative sizes of the components, their shapes, spins, and thermal and mechanical properties \cite{walsh_AIV}. In binaries with unequal components (size ratio $<$0.6, called Group A in Fig.~3), the larger (primary) body retains the fast spin rate of the precursor and only the secondary can be synchronised with the binary orbit \cite{jacobson-scheeres2011}. In binaries with a larger size ratio (Group B) the component spin rates and binary orbital period can be synchronised by mutual tides. Binary systems created directly from the rotational fission of a strengthless precursor body can have semimajor axes of up to 34$R_p$, where $R_p$ is the radius of the primary \cite{jacobson-scheeres2011}. The 288P system has a semimajor axis of at least 76$R_p$, and so cannot have formed directly from rotational fission of a strengthless precursor. 
The semimajor axis of a tidally locked binary system can, however, be expanded beyond the 34$R_p$ limit through the action of radiative torques (binary YORP or BYORP effect) \cite{cuk-burns2005}. At least in systems with a low size ratio (Group A), this can lead to the formation of Wide Asynchronous Binaries (Group W), which remain stable after the secondary spin and orbital period decouple \cite{jacobson-scheeres2014}. 

Wide binaries might also form in the aftermath of a catastrophic impact generating fragments of similar size that subsequently enter into orbit about each other \cite{durda-bottke2004a}. It is possible that the event forming the (7.5 $\pm$ 0.3) million year old 288P family \cite{novakovic-hsieh2012} created such an Escaping Ejecta Binary (EEB). EEBs contain $<$10\% of the total mass involved in a catastrophic collision \cite{durda1996,durda-bottke2004a}, such that they are less numerous than single fragments susceptible to rotational splitting. If formed as an EEB, the activity of 288P might have been triggered by a more recent sub-catastrophic impact or rotational mass-shedding following YORP-spin up of one of the components not causally related to the binary formation. The average time interval between impacts of the relevant size is 10$^5$ years (see Methods), and the YORP spin-up timescale of a 1\,km asteroid is 10$^5$-10$^6$ years \cite{jacobson-scheeres2011} with a variation of orders of magnitude because the YORP effect depends sensitively on a body's shape and material properties. Hence, both impact activation and YORP-driven rotational fission are plausible in the time since the family-forming collision.

The high eccentricity of the system is consistent with both the EEB and the rotational fission scenario, as tidal damping of the eccentricity occurs on timescales longer than the age of the 288P family \cite{jacobson-scheeres2014}.

The YORP torque influences the obliquity \cite{vokrouhlicky-nesvorny2003,hanus-durech2011,cibulkova-durech2016}, driving about 50\% of the objects to obliqities of 0$^\circ$ or 180$^\circ$ \cite{vokrouhlicky-capek2002}. Therefore, the mutual orbit of a binary system formed by rotational fission has an elevated probability to be aligned with the heliocentric orbit, as 
is observed in 288P (see Extended Data Figure 1). If 288P were an EEB, the alignment of the binary and heliocentric orbits would have to be considered a coincidence for which the statistical probability is $\sim$1\% (see Extended Data Figure 1).
Given this low probability and the low mass fraction of EEBs indicated by collision models, rotational fission seems the more likely formation process of 288P. 

Surface ice cannot survive in the asteroid belt for the age of the solar system but can be protected for billion-year timescales by a refractory dust mantle only a few meters thick \cite{schorghofer2008}. It is therefore likely that an event splitting a body into two parts of similar size will uncover buried ice if present. A decisive factor for the subsequent development of the system is whether the sublimation will last longer than the time required to tidally synchronise the spin and binary orbital periods, which is 5,000 years for equal-mass components but orders of magnitude longer for lower mass ratios \cite{jacobson-scheeres2014}. Sublimation-driven activity can last longer than 5,000\,years \cite{capria-marchi2012}, such that for high-mass ratio systems it is conceivable that activity prevails after tidal synchronisation. 
In this case, the recoil force from the local sublimation of water ice can drive binary evolution. Subject to the many unknowns, we find that the timescale to change the orbit of a synchronous binary system by sublimation torques can be several orders of magnitude shorter than for radiation torques (see Methods).
For this reason it seems more likely that 288P's wide separation reflects the action of sublimation torques, although BYORP and subsequent re-activation cannot be excluded. The discussed evolutionary paths are illustrated in Extended Data Figure 5.

Most asteroid binaries are discovered either by radar, when close to the Earth, or by mutual eclipses in their lightcurves, when the component separations are small. Kilometer-sized asteroids in the main-belt are too small and distant to be studied by radar, while wide binaries align to produce mutual eclipses only rarely. As a result, there is a very strong observational bias against the detection of small, wide main-belt binaries of the sort exemplified by 288P. The binary nature of 288P was discovered as a by-product of the activity of this body, which attracted attention and motivated the initial HST observations. While there are many biases against the detection of wide binaries in the asteroid belt, there is no obvious bias against detecting systems with similar component sizes. Still, the previously known six wide binaries have a diameter ratio $\sim$0.3 (Fig.~3) whereas in 288P this ratio is close to unity. This suggests that 288P is of a rare type even beyond the detection bias. A larger sample of wide binaries is needed to establish whether high-mass ratio systems are more likely to be active than low-mass ratio systems. Based on currently available models, the most probable formation scenario of 288P is rotational breakup followed by rapid synchronisation and orbit extension by sublimation torques. This path would be much less probable in low-mass ratio systems due to the longer synchronisation timescale. It is therefore possible that the activity played a decisive role in the formation of the 288P system, and that the high mass ratio was a prerequisite for that.


\noindent{\bf Acknowledgements.} This work is based on observations made with the NASA/ESA Hubble Space Telescope, obtained at the Space Telescope Science Institute, which is operated by the Association of Universities for Research in Astronomy, Inc., under NASA contract NAS 5-26555. These observations are associated with programs \#12597, \#14790, \#14864, and \#14884.\\

\noindent{\bf Author contributions.} J.A. identified the potential binary nature of 288P, applied for HST observing time, carried out the model calculations regarding the binary orbit and the dust dynamics, and led the effort preparing the manuscript. D.J. calculated the importance of the sublimation-driven torque and contributed to the interpretation and presentation of the data. M.M. processed the raw images and was responsible for the removal of cosmic rays and the production of the sub-sampled composite images. H.W. contributed to designing and preparing the observations. S.L. checked the work and critiqued the proposals and paper.\\

\noindent Correspondence and requests for materials should be addressed to agarwal@mps.mpg.de\\

\noindent The authors declare no competing financial interests.\\

\clearpage

\noindent\underline{\bf\Large Methods}\\

\noindent{\bf Orbit calculation.}
The relative motion of two bodies in orbit about their  centre of mass can be described by a Keplerian ellipse with one of the bodies fixed in one focus point and the other orbiting it along the periphery according to Kepler's laws. The length of the radius vector of the ellipse corresponds to the objects' mutual distance, and the true anomaly to the angular distance from the common semimajor axis of the system. The eccentricity and period are the same as for the two individual orbits.

The line connecting the two nuclei is in all images consistent with the projected orbit, and the angle between the line of sight from the Earth to 288P and its orbital plane was during all observations $<$2.3$^\circ$. We therefore assume for the following model that the observer was always in the orbital plane of the binary system. 

Extended Data Figure~2a shows the relative orbit of the binary system and a line of sight from Earth, as they would be seen from an ecliptic northern polar position. The apparent physical distance, $d$, of the components at the time $t$ is described by $d(t) = |sin (\theta_p(t) - \alpha(t)|$, where $\theta_p(t)$ is the true anomaly for a prograde orbit, and $\alpha(t)$ is the angle between the system's semimajor axis and the line of sight. For a retrograde orbit, and keeping the definition of $\alpha$, the distance is given by $d(t) = |sin (\theta_r(t) + \alpha(t)|$.

The angle $\alpha$ changes with time due to the relative motion of the Earth and the binary system. Extended Data Figure~2b shows the apparent motion of 288P during the time frame of our observations in the observer-centred ecliptic coordinate system. While the ecliptic longitude varies by 25$^\circ$, the latitude changes by only 3$^\circ$. We therefore approximate the change in $\alpha$ by the change in observer-centred ecliptic longitude $\lambda$. We define $\alpha_0$ to be the angle between the line of sight and the system's semimajor axis during the first HST observation on 2016 August 22, and $\alpha_0$ is a free parameter of our orbit-fitting simulation. The time-dependence of $\alpha$ is then given from the known change in $\lambda$, with $\alpha(t) = \alpha_0 + \lambda(t) - \lambda_0$, where $\lambda_0$ is the observer-centred ecliptic longitude of 288P on 2016 August 22.\\

\noindent{\bf System Mass and Density.}
The density is calculated from the total mass, $M$, and volume $V$ of the system. The mass is given by Kepler's law
\begin{equation}
M = \frac{4 \pi^2 a^3}{G P^2},
\label{eq:mass}
\end{equation}
where $G$ is the gravitational constant and $P$ is the orbital period and is found to be in the range $1.3 \times 10^{12} < M < 1.1 \times 10^{13}$ kg for the combinations of $a$ and $P$ compatible with the data (Fig. 2).
The total volume, $V$, of the two nuclei, is approximated by that of two spheres having the total cross-section $A$:
\begin{equation}
V = \sqrt{\frac{8}{9\pi}} A^{3/2},
\label{eq:volume}
\end{equation}
Assuming $A$=5.3$\times$10$^6$\,m$^2$ \cite{agarwal-jewitt2016}, we find $V$=6.5$\times$10$^9$\,m$^3$. To estimate the uncertainty of the volume, we consider the ratio of the smallest to the largest observed cross-section for one of the components to be 0.7, corresponding to a lightcurve amplitude of 0.4\,mag \cite{waniak_288P_dps}. Not knowing at which rotational phase our observation was made, we estimate that our measured cross-section represents the mean cross-section with an uncertainty of 20\%, and that therefore the uncertainty of the volume estimate is 30\%. This is a lower limit, because we do not know the extent of the components in the third dimension and the overall shapes of the bodies. To account for these and the (comparatively small) uncertainty of the albedo, we assume a total volume uncertainty of 60\%. 
Combining the smallest (largest) possible mass with the largest (smallest) possible volume, we find densities between 120\,kg\,m$^3$ and 4200\,kg\,m$^3$, consistent with typical asteroid densities of 1500\,kg\,m$^{-3}$ \cite{hanus-viikinkoski2017}.\\

\noindent{\bf Dust production.} 
We estimate the dust production rate from the brightness of the coma within a projected aperture of 400\,km (corresponding to between 8 and 15 pixels, depending on geocentric distance). For each observation, we measured the flux $F_{ap}$ within circular apertures of increasing radius $r_{ap}$. The flux rises linearly with $r_{ap}$, with different slopes for $r_{ap}<$7\,px and  $r_{ap}>$7\,px. Assuming that at $r_{ap}>$7\,px, the surface brightness is dominated by dust, we fit a linear relation $F(r_{ap}) = F_n + k r_{ap}$ to $F_{ap} (r_{ap})$, where $F_n$ is the nucleus flux and $F_c(r_{ap}) = k r_{ap}$ is the flux of light reflected by dust inside the aperture. The uncertainty of the flux measurement is small compared to those of the albedo, phase function, bulk density, size, and velocity of the dust used in the following to convert the surface brightness to a production rate.

We convert the measured flux $F$ (in electrons/s) to apparent magnitudes using $m_V = -2.5 \log10 F + Z$, with $Z$=25.99 for the F606W filter \cite{kalirai-mackenty2009}, and to absolute magnitudes $H_V$ assuming a C-type phase function with $G$=0.15. Using instead an S-type phase function with $G$=0.25 would render $H_V$ fainter by 0.14\,mag at the largest observed phase angle, reducing the corresponding dust cross-section by 10\%.
 
The total dust cross-section in the aperture is given by $C = 1329^2 \pi 10^{-0.4 H_V} / (4 p_V)$, where we use a low geometric albedo of $p_V$=0.05. With $p_v$=0.1, the dust cross-section would reduce by a factor 2. Our employed combination of $G$=0.15 and $p_V$=0.05 implies that the derived cross-section is at the lower end of the possible range.

We convert this area to a mass assuming representative particle radii of 6 and 60\,$\mu$m, respectively, and a bulk density of 1000\,kg\,m$^{-3}$, which is also a low value, with typical C-type nucleus densities ranging from 1000 to 2000\,kg\,m$^{-3}$ \cite{hanus--viikinkoski2017}, such that the derived mass represents a lower limit and could be a factor 4 higher. Additional uncertainty is introduced by our lack of knowledge if the density of asteroid dust can be compared to that of the nuclei, and if dust of the same size dominates the optical cross-section and the mass of the ejected material.

Using the velocity-size relation derived from 2011 HST data \cite{agarwal-jewitt2016}, we calculate the dust production rate from the time that a dust particle would remain inside the aperture depending on its size. The statictical uncertainty of the velocity is 30\% (from the scatter of the data points in Figure 11 of Reference \cite{agarwal-jewitt2016}). This velocity represents a lower limit because it is only the component perpendicular to the orbital plane, such that also the derived production rate is a lower limit. 
Extended Data Fig.~4 shows the inferred dust production rates for the two different assumptions of the dominant grain size. \\

\noindent{\bf Impact timescale.} We estimate the average time interval between impacts excavating the amount of ice required to explain the observed dust production as follows. To explain the dust production rate of 1\,kg\,s$^{-1}$ \cite{agarwal-jewitt2016}, and assuming a dust-to-gas mass ratio of 1 -- 10, an ice-sublimating active patch of (30 -- 90)\,m in radius is required on a perfectly absorbing body at the heliocentric distance of 2.45\,AU. A crater of this size on a strengthless rubble pile would have been generated by a 1\,m-sized projectile \cite{housen-holsapple2011} impacting at the typical relative velocity of main belt objects of 5\,km\,s$^{-1}$ \cite{bottke-nolan1994}. The collisional lifetime (probability to be impacted by a 30\,m radius asteroid) of a 1\,km radius main belt asteroid is 10$^9$ years \cite{bottke-durda2005b}. The abundance of 1\,m scale asteroids is uncertain, but they are probably a factor of $\sim$10$^4$ more numerous than those with 30\,m radius \cite{bottke-durda2005b}, such that the time interval between impacts of 1\,m bodies on a 1\,km asteroid is 10$^5$ years, considerably less than the age of the 288P family. Impact activation is therefore plausible. \\

\noindent{\bf Orbital torque by sublimation.} Assuming that the dust production was driven by a comparable gas production rate $Q_{gas}$, and that the gas was leaving the nucleus with the thermal expansion speed of $v_{th}$ from a small patch, this directed emission of gas exerts a torque, $T$, which can have influenced the binary orbit if the torque was tangential to the orbit, and the orbit and the rotation of the active component were synchronous. The maximum torque is given by 
\begin{equation}
T = k Q_{gas} v_{th} r,
\end{equation}
where 0$<k<$1 is a dimensionless parameter describing the degree of collimation of the gas flow (with $k$ = 0 corresponding to isotropic ejection and $k$ = 1 to perfectly collimated ejection), and $r$ is the radius vector of the binary orbit. Over one mutual orbit of period $P$, this gives a change in angular momentum of 
\begin{equation}
\Delta L = k  Q_{gas} v_{th} \int_0^P r dt.
\end{equation} 
We approximate this by $\Delta L=k  Q_{gas} v_{th} a P$, and assume $k$=0.1, $v_{th}$=500\,m\,s$^{-1}$, $Q_{gas}$ = 0.1\,kg\,s$^{-1}$, and an initial $a$=30\,km, and $P$=30\,days, obtaining $\Delta L$ = 4$\times$10$^{11}$\,kg\,m$^2$\,s$^{-1}$. Comparing this to the total angular orbital momentum of 288P ($\sim$5$\times$10$^{14}$\,kg\,m$^2$\,s$^{-1}$), and given that 288P is active for $\sim$10\% of each orbit, we find that it would take of order 10$^4$ revolutions of the binary orbit ($\sim 5\times10^3$\,years) to change the total angular momentum by a factor $\sim$2. We note that both the $k$-parameter and $Q_{gas}$ influence $\Delta L$ linearly, such that the timescale easily has an uncertainty of an order of magnitude or more. Nevertheless, the calculation shows that sublimation torques can change a binary orbit over much shorter timescales than the photon-driven BYORP-effect, which doubles the semimajor axis in (3-6)$\times$10$^4$\,years \cite{jacobson-scheeres2014}.

\noindent{\bf Code availability.} We have opted not to make the code used to calculate the orbit fit and the synchrone-syndyne analysis available because custom routines were developed for this analysis.\\

\noindent{\bf Data Availability.} The HST datasets analysed during the current study are available in the Mikulski Archive for Space Telescopes (https://archive.stsci.edu). The orbital data shown in Figure 3 are available in the NASA Planetary Data System under identifyer EAR-A-COMPIL-5-BINMP-V9.0 (https://pdsquery.jpl.nasa.gov). All other data sets generated during the current study are available from the corresponding author on reasonable request.
\\

\clearpage

\begin{figure}
\includegraphics[width=\textwidth]{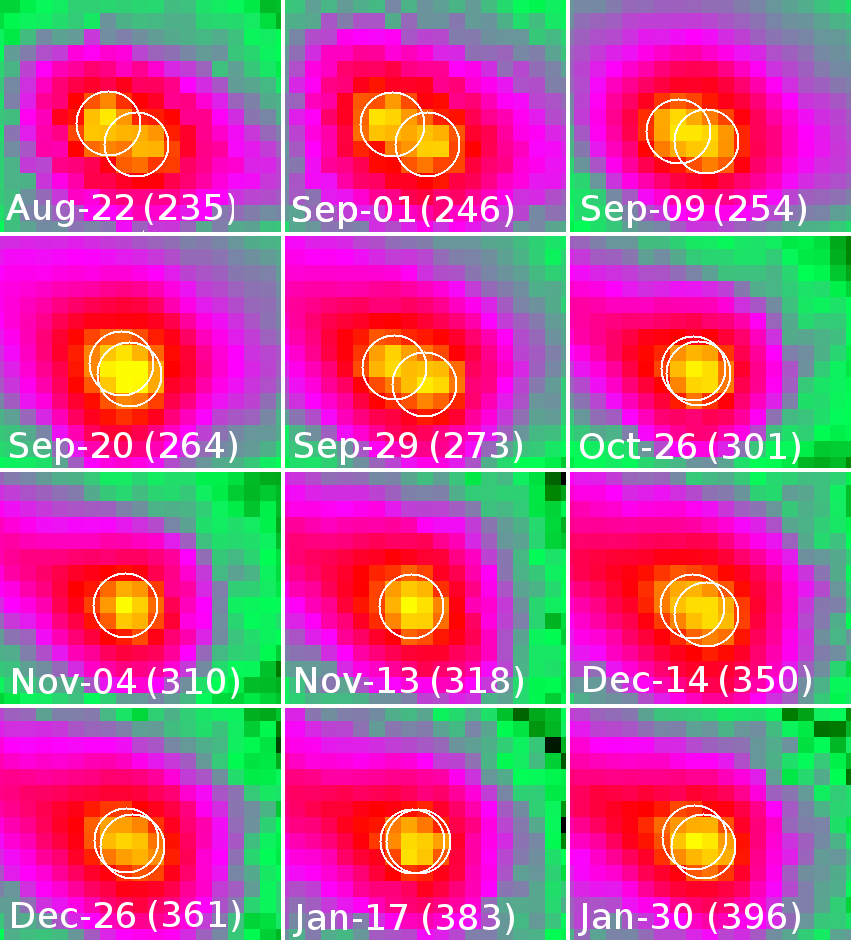}
\caption{The 288P system between August 2016 and January 2017.  The images were obtained with the 1k$\times$1k C1K1C subarray of the Wide Field Camera 3 of the HST and the wide passband filter F606W, centred at 595.6\,nm. Details of the observations are listed in Extended Data Table 1. Each panel is a composite of 8 single exposures of 230\,s, obtained with a 2$\times$2 sub-sampling dither-pattern that enabled us to re-sample the images to a pixel scale of 0.025\arcsec. Each panel is 4.5\arcsec$\times$3.8\arcsec\ in size. The intensity scale is logarithmic, and the range was adjusted manually for each image to account for the changing brightness. The appearance of 288P alternated between two clearly separated nuclei of similar brightness and a single point source, confirming that 288P is a binary asteroid. We measured the distance by visually fitting circles of 2 pixels radius to the point spread functions (PSFs) of the two components. We estimate the 3-sigma uncertainty of the measured distance between their centres to be $\pm$0.5\,pixel ($\pm$0.013\arcsec). The numbers in parentheses indicate the day of the year (DOY) 2016.}
\label{fig:nuclei}
\end{figure}

\begin{figure}
\centering
\includegraphics[width=0.7\textwidth]{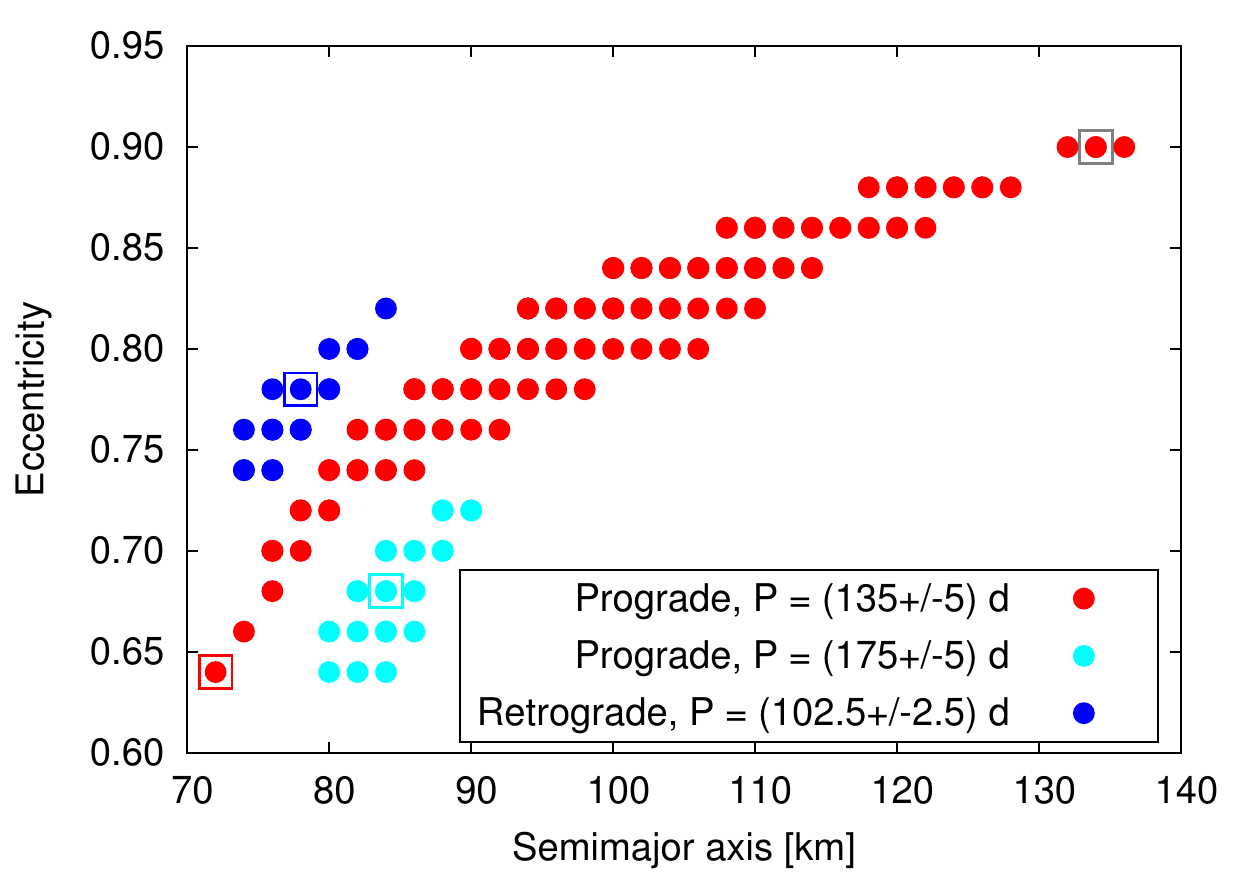}
\includegraphics[width=0.7\textwidth]{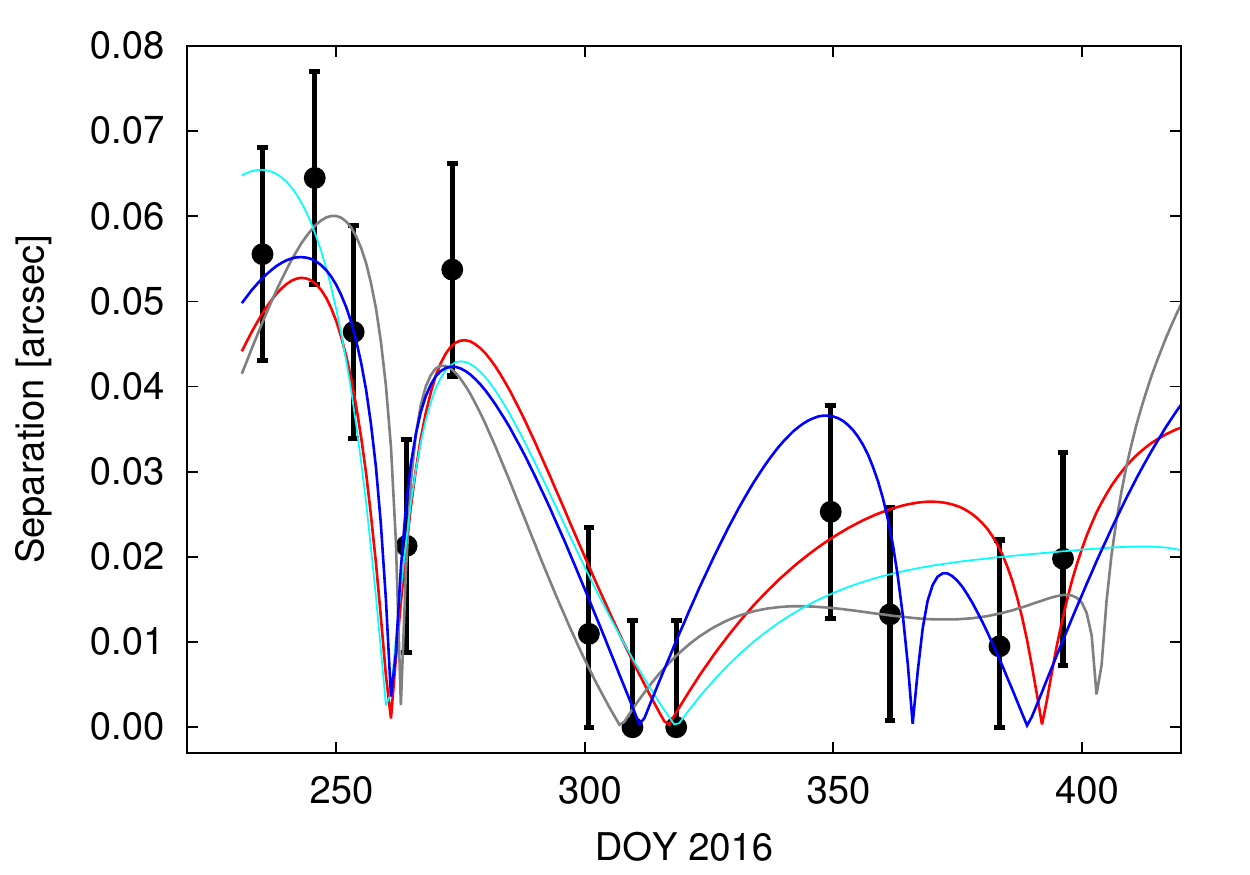}
\caption{Binary orbits matching the observations. To infer the Keplerian elements of the mutual orbit, we calculated the projected distances at the times of the observations for a large set of both prograde and retrograde orbits, varying 5 parameters independently: the semimajor axis, $a$, between 40 and 150 km in steps of 2 km, the eccentricity, $e$, between 0 and 0.98 in steps of 0.02, the orbital period, $P$, between 20 and 210 days in steps 5 days, the time of perihelion in steps of 1/20 of the orbital period, and the angle, $\alpha_0$, between the perihelion vector and the line of sight on 22 August in steps of 10 degrees. To account for the changing observing geometry, we subtracted the difference in geocentric ecliptic longitude between 22 August and the date of observation from $\alpha_0$ for each observation date (see Methods and Extended Data Figure 3). We searched this parameter space for combinations reproducing all 12 measurements. Panel a shows the acceptable combinations of the semimajor axis $a$ and eccentricity $e$.
Red and light blue symbols refer to prograde orbits with (130$<P<$140)\,days and (170$<P<$180)\,days, respectively, while dark blue symbols represent retrograde orbits with (100$<P<$105)\,days. All solutions have the line of sight on 22 August within $\pm$10$^\circ$ of the system's major axis, and a periapsis date between 16 and 21 September.
Panel b shows the measured and simulated component distance for 
four representative orbit solutions marked by boxes of the same colour in the upper panel. These four solutions were chosen to reflect the diversity of the possible orbits. The error bars of $\pm$0.013\arcsec\ reflect the estimated 3-sigma position uncertainty of the circles in Fig.~1. The measured component distances are listed in Extended Data Table 2.}
\label{fig:ae}
\end{figure}

\begin{figure}
\includegraphics[width=\textwidth]{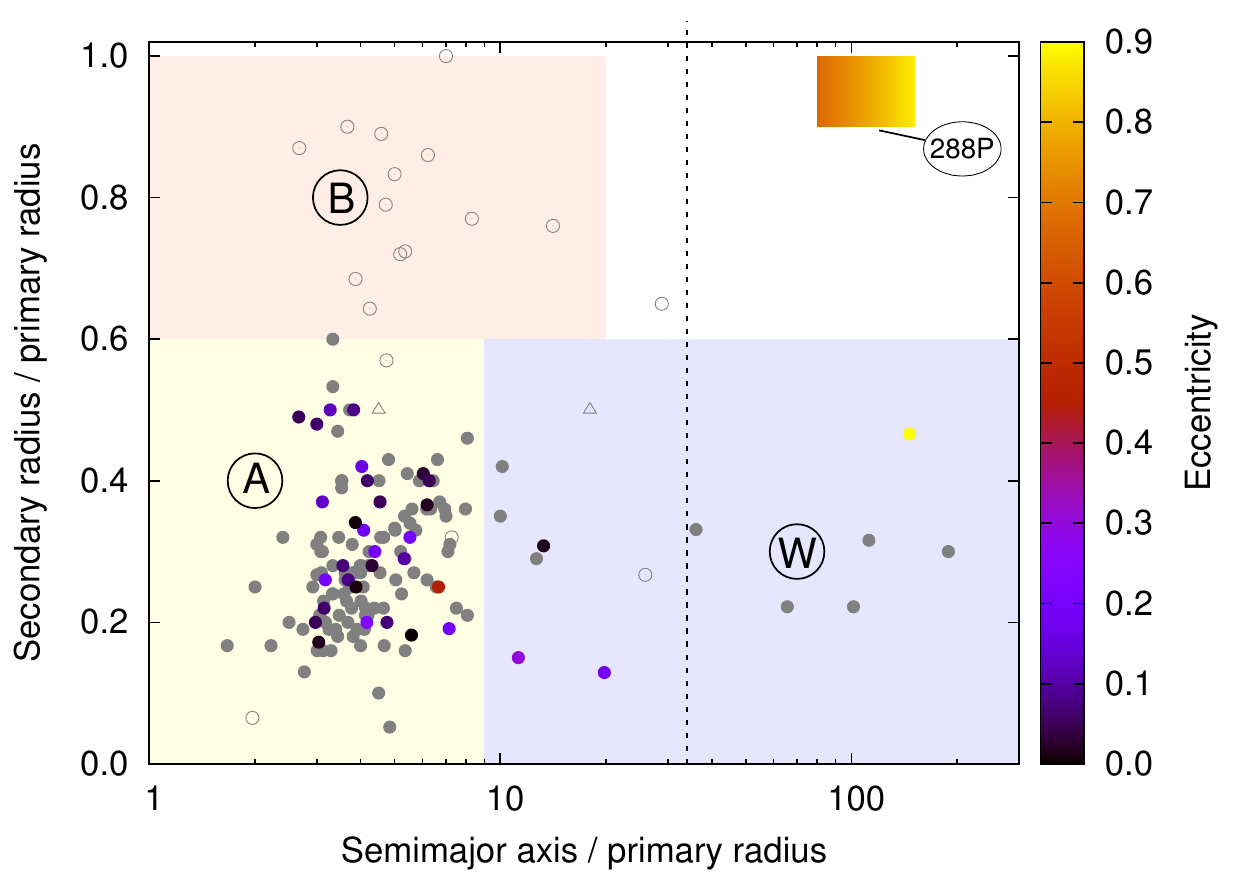}
\caption{Orbital properties of 288P and previously known binary asteroids. The plot shows the size ratio as a function of the semimajor-axis-to-primary-radius ratio for all asteroids with known primary and secondary radius and semimajor axis \cite{johnston2016}. The eccentricity is colour-coded, with grey symbols used for systems with unmeasured eccentricity. 
Filled circles represent systems with a primary rotation period $P<$5h, open circles indicate $P>$5h, and triangles an unknown primary rotation period. The dotted line corresponds to 34\,$R_p$, the upper limit for binaries to form directly from a strengthless precursor \cite{jacobson-scheeres2011}. 
The letters A, B, and W and the colour shading reflect the three major groups of known small asteroid binaries \cite{pravec-harris2007}. Group A binaries have a size ratio $<$0.6 and a fast rotating primary and, in 2/3 of the systems, a secondary rotating synchronously with the binary orbit. Group B consists of doubly synchronous systems with similar component size, and Group W consists of wide, asynchronous binaries. All three groups are consistent with an origin by rotational fission \cite{jacobson-scheeres2011}. The effect of tides on the spin state depends on the component size ratio and distinguishes Groups A and B. Group W possibly has evolved out of Group A under the action of the BYORP effect \cite{jacobson-scheeres2014}. 
288P occupies a region in this parameter space that has until now been unpopulated. We estimate a lower limit of 0.8 for its cross-section ratio from the 0.2\,mag maximum brightness difference of the two components in individual exposures. This corresponds to a radius ratio of 0.9. The combined double-peaked lightcurve of 288P shows a 16\,hour periodicity \cite{waniak_288P_dps}. This constrains the more variable component to a 16\,hour rotation period, while the rotation of the second component, if less variable, is not well constrained by the lightcurve.}
\label{fig:binaries}
\end{figure}

\clearpage
\noindent\underline{\bf\Large Extended Data}

\renewcommand{\figurename}{Extended Data Figure}
\renewcommand{\tablename}{Extended Data Table}
\setcounter{figure}{0}    
\setcounter{table}{0}

\begin{table}[h]
\caption{Parameters of the HST observations. $N$ is the sequence number of the observation, $r_h$ and $\Delta$ are the heliocentric and geocentric distances in AU, $\alpha$ is the phase angle, $PA_{-\odot}$ and $PA_{-v}$ are the position angle of the anti-solar direction and of the projected negative orbital velocity vector, $\epsilon$ is the angle between the line of sight and the orbital plane of 288P, and long and lat are the observer-centred ecliptic longitude and latitude.}
\vspace{2mm}
\centering
\label{tab:observations}
\begin{tabular}{llrrrrrrrrr}
\hline\noalign{\smallskip}
N & UT Date & DOY16 & $r_h$ & $\Delta$ & $\alpha$ & $PA_{-\odot}$ & $PA_{-v}$ & $\epsilon$ & long & lat\\
 &  &  & [AU] & [AU] & [deg] & [deg] & [deg] & [deg] & [deg] & [deg]\\
\hline\hline\noalign{\smallskip}
1 & 2016-Aug-22 & 235.16 & 2.47 & 1.50 &  8.95 & 259.53 & 246.55 &  2.00 & 351.19 &  5.23\\
2 & 2016-Sep-01 & 245.67 & 2.46 & 1.46 &  4.54 & 275.56 & 246.88 &  2.17 & 349.36 & -5.41\\
3 & 2016-Sep-09 & 253.50 & 2.45 & 1.45 &  2.25 & 330.82 & 247.19 &  2.24 & 347.81 & -5.46\\
4 & 2016-Sep-20 & 264.16 & 2.45 & 1.46 &  5.33 & 42.83 & 247.64 &  2.22 & 345.69 & -5.43\\
5 & 2016-Sep-29 & 273.33 & 2.44 & 1.49 &  9.23 & 54.45 & 248.00 &  2.12 & 344.13 & -5.32\\
6 & 2016-Oct-26 & 300.83 & 2.44 & 1.69 & 18.65 & 63.63 & 248.43 &  1.45 & 342.31 & -4.63\\
7 & 2016-Nov-04 & 309.60 & 2.44 & 1.78 & 20.61 & 64.73 & 248.31 &  1.18 & 342.83 & -4.36\\
8 & 2016-Nov-13 & 318.42 & 2.44 & 1.88 & 22.06 & 65.50 & 248.08 &  0.90 & 343.87 & -4.10\\
9 & 2016-Dec-14 & 349.50 & 2.44 & 2.27 & 23.74 & 66.83 & 246.86 &  0.01 & 351.00 & -3.25\\
10 & 2016-Dec-26 & 361.40 & 2.45 & 2.42 & 23.29 & 67.12 & 246.40 & -0.27 & 354.83 & -2.97\\
11 & 2017-Jan-17 & 383.46 & 2.46 & 2.70 & 21.35 & 67.74 & 245.83 & -0.66 &   2.92 & -2.54\\
12 & 2017-Jan-30 & 396.23 & 2.47 & 2.85 & 19.70 & 68.26 & 245.74 & -0.82 &   8.16 & -2.31\\
\noalign{\smallskip}\hline
\end{tabular}
\end{table}

\clearpage

\begin{table}[h]
\caption{Measured component separations $S$ from Fig.~1. For observations with separations $>$2\,pixels (0.05\arcsec), the brightness of the individual components is also listed, where $F_E$ and $F_W$ refer to the Eastern and Western component, respectively. The values represent the total flux within an aperture of radius 1.5\,pixels ($r_{ap}$=0.0375\arcsec) centred as indicated by the circles in Fig.~1 and are not background-subtracted due to the unknown distribution of the dust. The point spread function (PSF) of WFC3/UVIS at 600\,nm is 0.067\arcsec\ \cite{dressel2017}, such that even at the largest observed separation, the PSFs of the two nuclei overlap. Each 0.0375\arcsec aperture encircles 90\% of the flux from the central nucleus. The energy from the neighbouring nucleus is contained to 83-88\% (for 0.054$<S<$0.065) within a circle of radius $S-r_{ap}$ not overlapping with the aperture and to 5\% outside a circle of radius $S+r_{ap}$ also not overlapping. Assuming that not more than half of the remaining energy falls into the aperture, this would be 3.5 - 6\% of the total energy from the neighbouring source. Disregarding the dust contribution, the similar flux measured in the two apertures therefore reflects a similar brightness of the two point sources.}
\vspace{2mm}
\centering
\label{tab:distances}
\begin{tabular}{lrrr}
\hline\noalign{\smallskip}
UT Date & $S$[\arcsec] & $F_E$[e$^-$/s] & $F_W$[e$^-$/s]\\
\hline\hline\noalign{\smallskip}
2016-Aug-22 & 0.0556 & 54.82 & 50.62\\
2016-Sep-01 & 0.0645 & 53.93 & 55.64\\
2016-Sep-09 & 0.0464 &&\\
2016-Sep-20 & 0.0213 &&\\
2016-Sep-29 & 0.0537 & 64.75 & 66.48\\
2016-Oct-26 & 0.0110 &&\\
2016-Nov-04 & 0.0000 &&\\
2016-Nov-13 & 0.0000 &&\\
2016-Dec-14 & 0.0253 &&\\
2016-Dec-26 & 0.0133 &&\\
2017-Jan-17 & 0.0095 &&\\
2017-Jan-30 & 0.0198 &&\\
\noalign{\smallskip}\hline
\end{tabular}
\end{table}

\clearpage

\begin{figure}[h]
\centering
\includegraphics[width=0.7\textwidth]{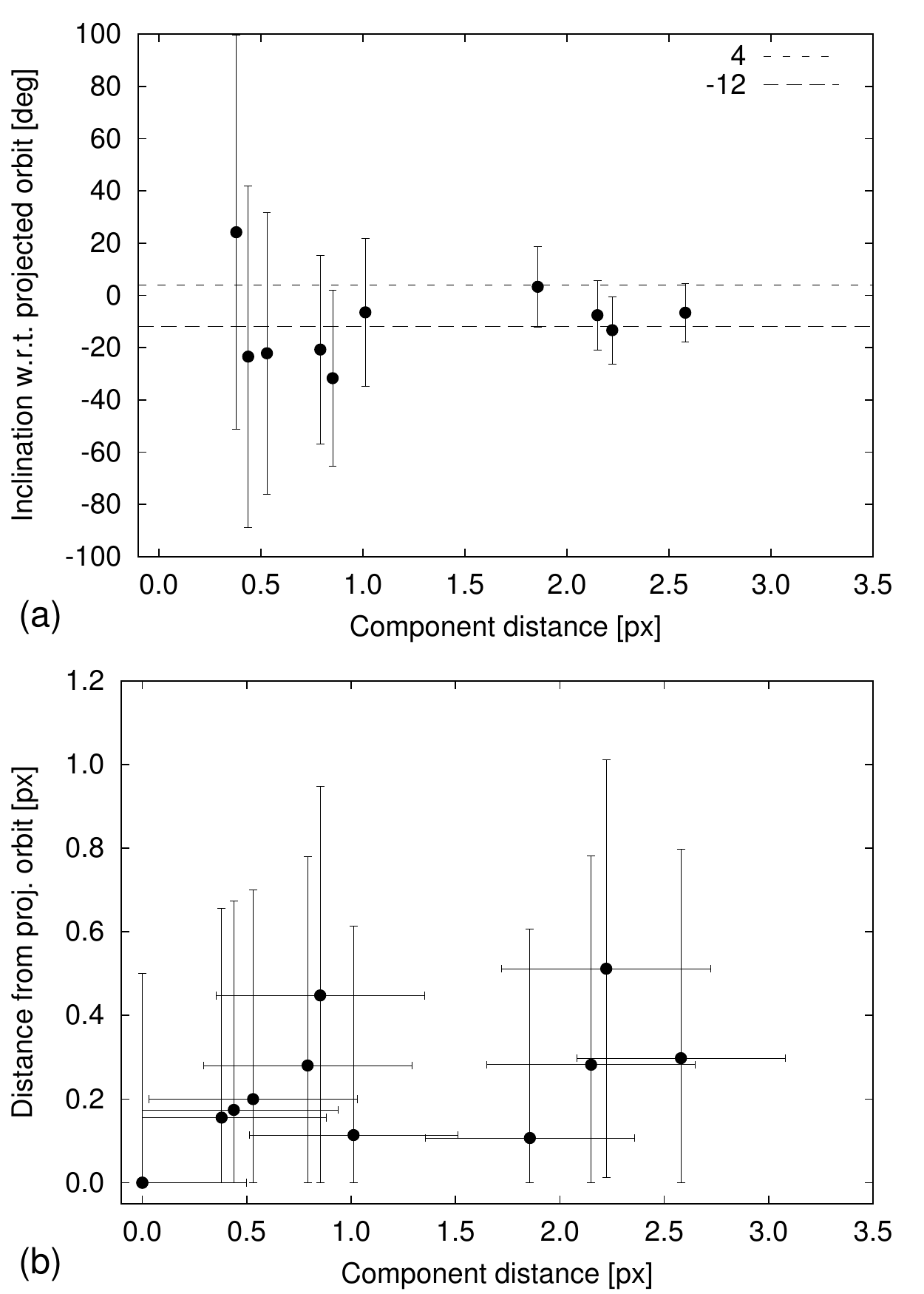}
\caption{Comparison of the binary orbit to the projected heliocentric orbit. Panel a shows the difference in on-sky position angle between the line connecting the two components and the projected heliocentric orbit. The measurements at large component distance ($>$1.5\,px) are consistent with projected inclinations between +4$^\circ$ and -12$^\circ$. The error bars in both panels represent the uncertainty propagated from the position uncertainty in Figure~1. Panel b shows the component distance perpendicular to the projected orbit, $\beta$. Near conjunction (separation $<$1.5\,px), these measure the angle $\alpha$ between the heliocentric and binary orbit perpendicular to the image plane through the relation $\sin \alpha = \Delta / D \sin \beta$, where $\Delta$ is the geocentric distance and $D$ is the component separation along the line of sight. We assume $D$=100\,km, and $\Delta$=2\,AU. With $\beta_{max}$=0.45\,px, we obtain $\alpha_{max}$=9$^\circ$. In conclusion, our best estimate of the binary orbit pole orientation is (-4$\pm$8)$^\circ$ in the image plane and (0$\pm$9)$^\circ$ perpendicular to it, and we describe the uncertainty of the pole direction by a double cone of opening angle 18$^\circ$. This corresponds to a solid angle of 0.15\,sr, or 1\% of 4$\pi$.}
\label{fig:plane}
\end{figure}

\clearpage 

\begin{figure}
\centering
\includegraphics[width=.7\textwidth]{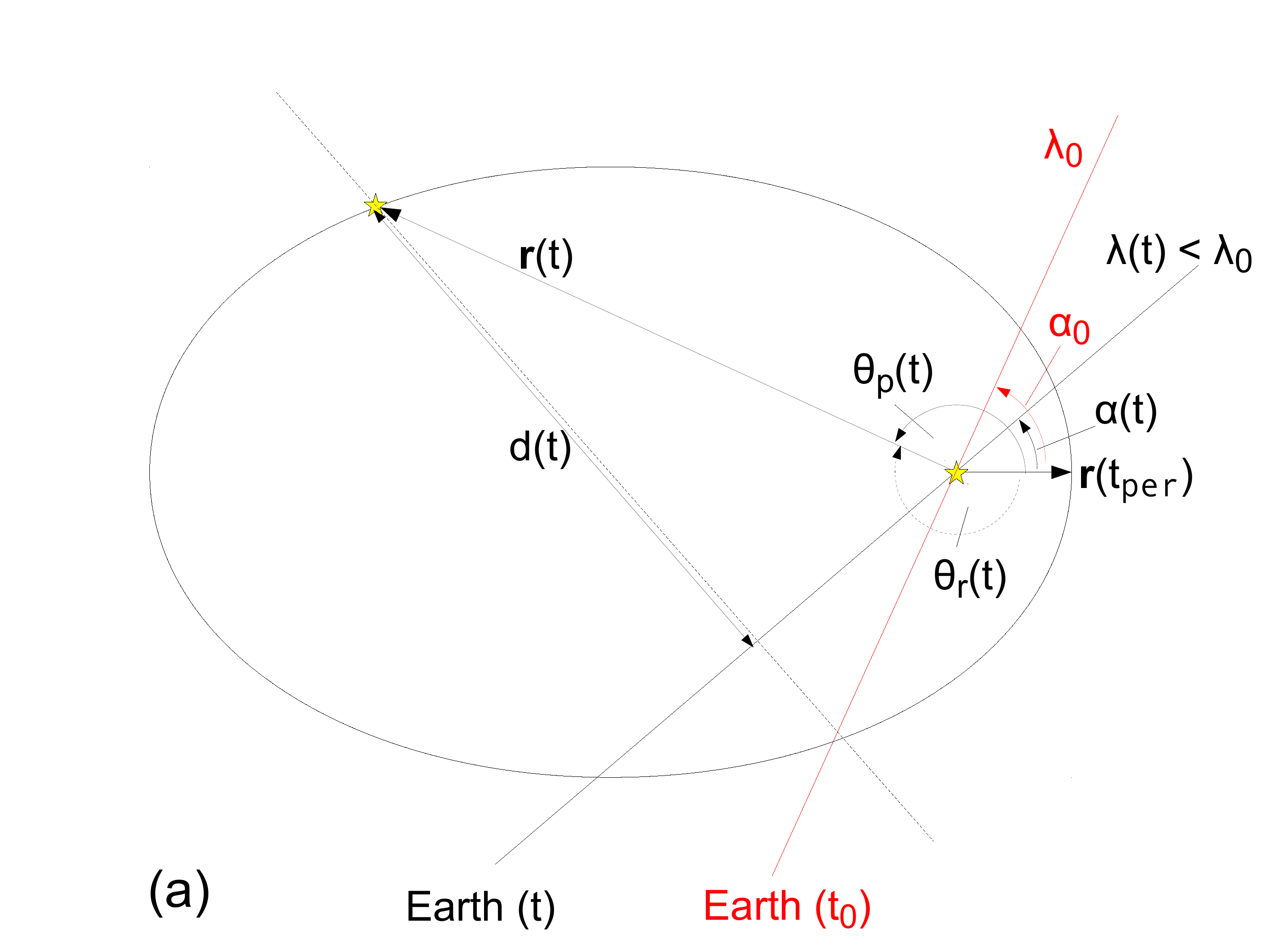}
\includegraphics[width=0.7\textwidth]{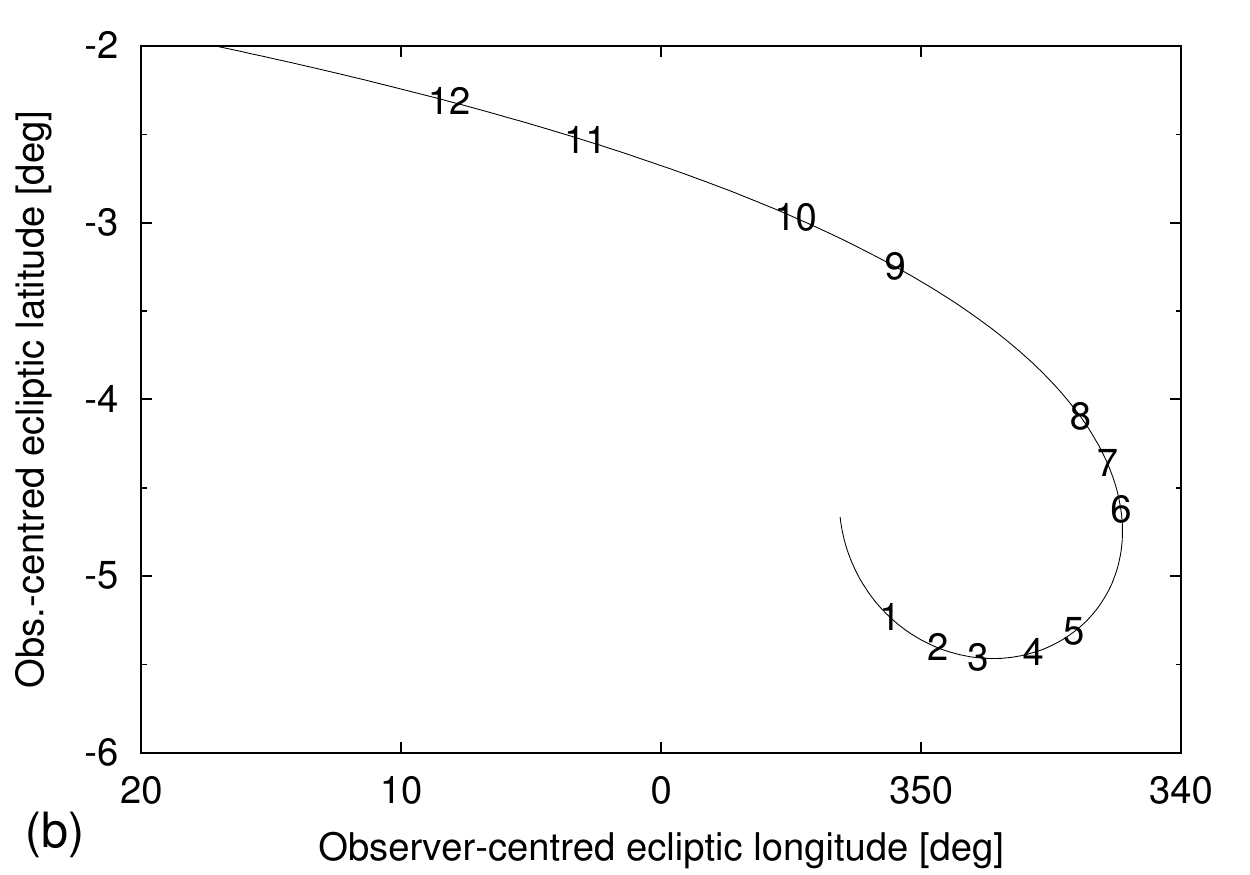}
\caption{Orbital and observational geometry during the HST observations.
a) Binary orbit and line of sight from Earth at an arbitrary fixed time $t$ (black) and with respect to the viewing geometry at a specific reference time (red), seen from the north ecliptic pole. The vector {\bf r}(t) describes the motion of one component with respect to the other fixed in one focus of the elliptic orbit. $t_{per}$: time of periapsis passage; $\theta_p, \theta_r$: true anomaly of a prograde and retrograde orbit; $d$: projected physical distance of the components; $\alpha (t)$: angle between the line of sight and the semimajor axis of the system; $\lambda$: observer-centred ecliptic longitude; the index 0 refers to the time $t_0$ (2016 August 22). b) Apparent motion of the 288P system to an Earth-based observer in ecliptic longitude and latitude over the timeframe of the HST observations. The coordinates at the times of the twelve observations are indicated by numbers, with 1 corresponding to 2016 August 22, and 12 to 2017 January 30 (see Extended Data Table~1).}
\label{fig:orbit_layout}
\end{figure}

\clearpage 

\begin{figure}
\includegraphics[width=\textwidth]{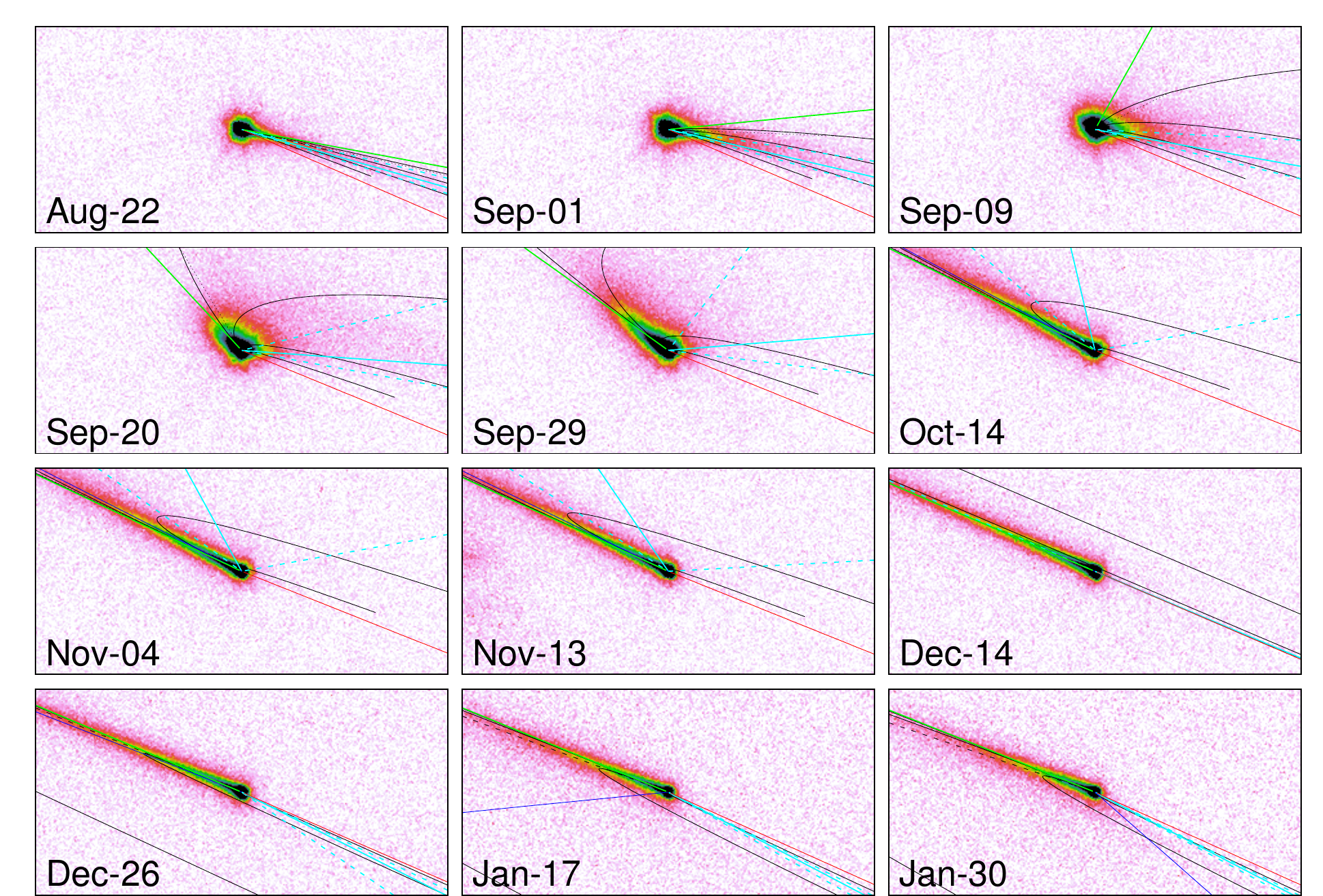}
\caption{The central 8\arcsec$\times$4\arcsec\ of the coma and tail of 288P. The red and green lines correspond to the projected orbit and projected anti-solar direction. Solid black lines show the loci of particles of fixed radiation pressure coefficient $\beta$ (syndyes \cite{finson-probstein1968a}), with $\beta$=10$^{-4}$,10$^{-3}$,10$^{-2}$,10$^{-1}$ in counterclockwise order. For a bulk density of 1000\,kg\,m$^-3$, this translates to particle sizes of 6\,mm, 600\,$\mu$m, 60\,$\mu$m, and 6\,$\mu$m, respectively. The remaining lines (cyan, blue, and black-dashed) show synchrones \cite{finson-probstein1968a}, the loci of particles ejected at a given time. The colours correspond to the following ejection dates: solid cyan: 2016 July 19, dashed cyan: 20 days before and after that date, blue: 2016 September 29, dotted black: 10 days before the observation, dashed black: 60 days before the observation. Up to September 09, the dust tail was oriented towards the direction where large (0.6 -- 6\,mm radius) dust grains ejected in July 2016 are expected. Beginning from September 20, a tail of 6 -- 60\,$\,\mu$m sized particles developed in the projected anti-solar direction, and remained there up to the end of our observation campaign in January 2017. On those dates when the viewing geometry allowed us to distinguish between 6 and 60\,$\,\mu$m (20 September to 26 October), the smaller size syndynes match the data better.}
\label{fig:synsyn}
\end{figure}

\clearpage
\begin{figure}[h]
\centering
\includegraphics[width=0.7\textwidth]{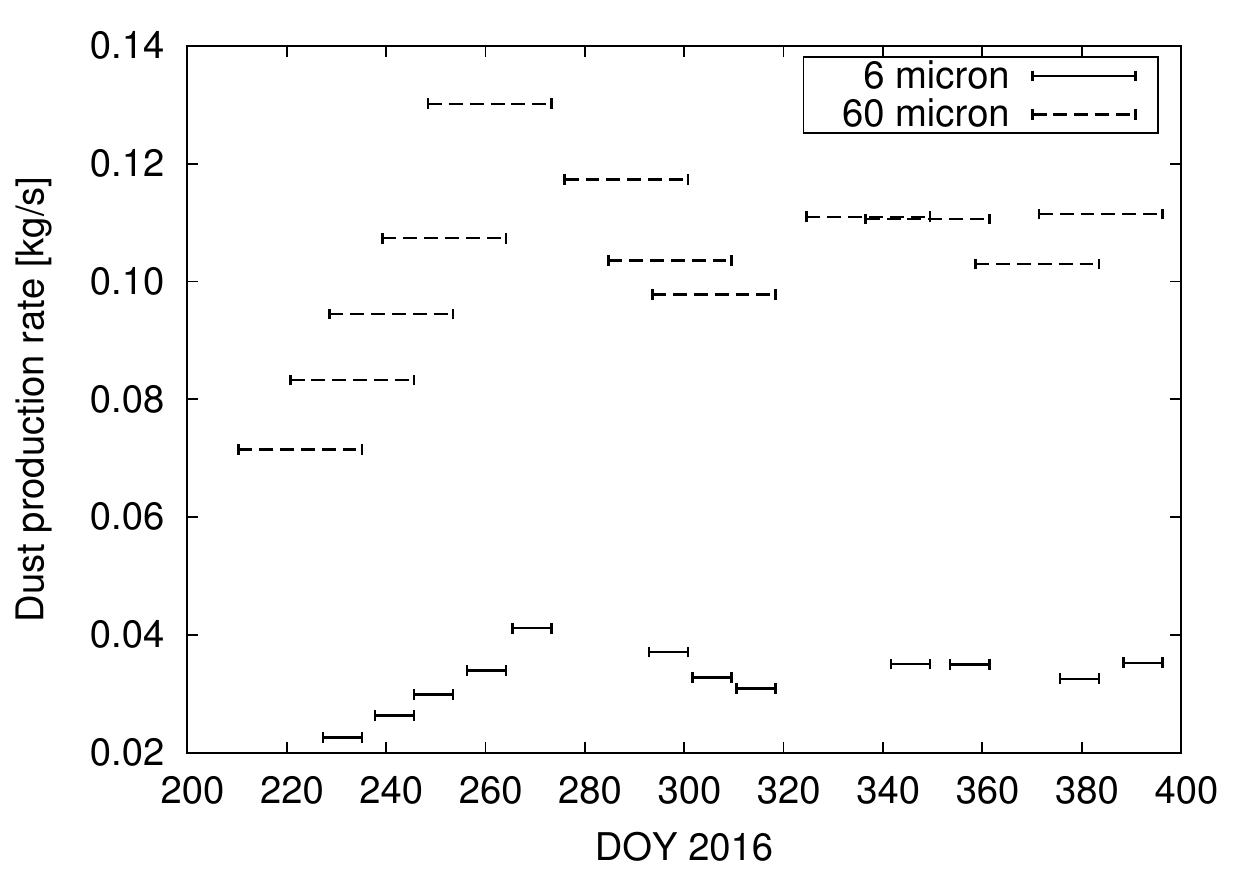}
\caption{Dust production of 288P. The production rate was inferred from the coma brightness within a 400\,km aperture for representative particle sizes of 6\,$\mu$m and 60\,$\mu$m. The production rates represent lower limits (see Methods). The horizontal error bars represent the time that it takes dust to leave the 400\,km aperture in which the dust brightness was measured.}
\label{fig:dp}
\end{figure}

\clearpage

\begin{figure}
\includegraphics[width=\textwidth]{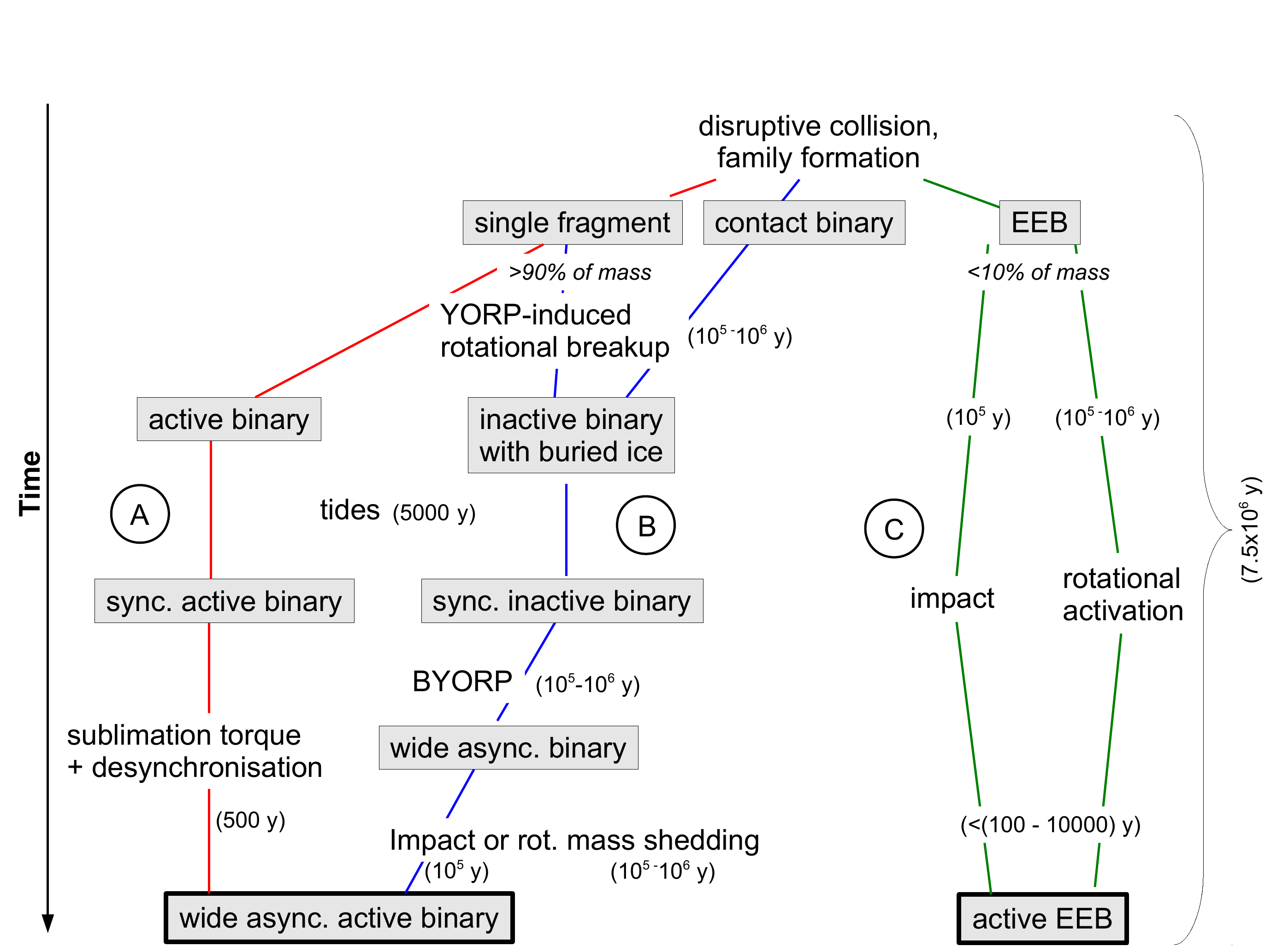}
\caption{Possible evolutionary paths of the 288P system. We assume that 288P is a fragment from a catastrophic collision 7.5$\times$10$^6$ years ago \cite{novakovic-hsieh2012}.
Possible outcomes of this collision are 1) a single fragment or a contact binary, or 2) an Escaping Ejecta Binary (EEB) \cite{durda-bottke2004a}. EEBs contain only a small fraction of the mass involved in a collision, while the bulk is in single fragments or contact binaries \cite{durda1996,durda-bottke2004a}. An EEB could subsequently have been activated by either an impact of a 1\,m radius body, or by rotational mass shedding after YORP-acceleration (path C). The average time between such impacts is 10$^5$ years, while the YORP spin-up time is 10$^5$-10$^6$ years \cite{jacobson-scheeres2011}. The sublimation can last between 100 and $>$5,000 years \cite{capria-marchi2012}. 
If 288P evolved out of a single fragment or a contact binary, it can have split into a binary by rotational fission on a timescale of 10$^5$-10$^6$ years. Subsequently, the binary and spin periods must have tidally synchronised, to enable BYORP or sublimation torques to further expand the semimajor axis. The timescale for tidal synchronisation of an equal-mass binary is 5,000 years \cite{jacobson-scheeres2011}, such that activity triggered upon splitting can have prevailed at the time of synchronisation. In that case (path A), sublimation torques can have expanded the binary orbit to its present state on timescales of 500 years. 
If the system was not active at the time of synchronisation (path B), the orbit expansion would have to be attributed to the BYORP effect, which takes several orders of magnitude longer than sublimation torques. The activity would in this case have had to be triggered by an impact or rotational mass shedding following renewed YORP spin-up. The timescales for path B are longer than for path A but well within the age of the 288P family.}
\label{fig:formation}
\end{figure}

\end{document}